\documentclass[a4paper]{panl}
\usepackage{cite}
\usepackage{wrapfig}
\usepackage{graphicx}
\usepackage{amssymb}
\usepackage{amsfonts}
\usepackage{amsmath}
\usepackage{longtable}
\usepackage{rotating}
\usepackage{lscape}
\usepackage{epsfig}
\usepackage{multirow}

\usepackage{hyperref}

\originalTeX
\begin{document}

\title{Fundamental physics asks philosophers new questions}
\maketitle
\authors{A.\,Panov\footnote{\label{footnote:email1}E-mail: panov@dec1.sinp.msu.ru}}
\from{M.V.Lomonosov Moscow State University, Skobeltsyn Institute of Nuclear Physics (SINP MSU), 1(2), Leninskie gory, GSP-1, Moscow 119991, Russian Federation}

\begin{abstract}
%
Modern fundamental physics poses new questions for philosophy, which, as it seems to us, have not yet received appropriate attention from philosophers of science. This paper formulates a number of such questions in order to present them to the attention, first of all, of professional philosophers. A rough list of the main themes is as follows: 1) Cosmic variance problem and the meaning of theoretical cosmology; 2) Epistemological status of the concept of Multiverse in cosmology; 3) The operational status of quantum macrostates and the relation of this problem to cosmology; 4) The meaning of the physical reality in the ``final theory''; 5) Criticism of the string theory in the relation with the item 4 above.
\end{abstract}
\vspace*{6pt}


\section*{Introduction}
\label{INTRODUCTION}

There  were a number of important breakthroughs of the last 40--45 years in fundamental physics. We mention here the following ones:

$\bullet$ Cosmology has become a precise quantitative science due to the emergence of inflation models, the associated theory of cosmological perturbations, and precision experimental data capable of verifying these models.

$\bullet$ The development of quantum information science, which makes intensive use of entangled quantum states and decoherence theory.

$\bullet$ The emergence of a large number of meaningful theoretical models of quantum gravity, including string theory, loop quantum gravity with spin foam models, etc.

Several new methodological issues have arisen in the depths of these new directions that have not been given sufficient attention by the philosophy of science. The question even arises: are all these issues really noticed by philosophers? A few of these problems without trying to solve them will be formulated here. We hope that this short report will help draw attention to these issues.

Before we begin the review, we will fix our understanding of the operational meaning of quantum mechanics and the objective nature of mathematical forms. This issues will be useful for later references.

\section{About the notions and meanings of experiment, measurement  and state in quantum physics (the ensemble approach)}
\label{OperationalQM}

Theoretical physics produces models that provide observable predictions. Experiments test the predictions of the theories, confirming the theories or disproving them. Verification of theories by experiment allows to improve the theoretical models or to create completely new theories, and this cycle is repeated again and again. It is the well-known cycle of empirical science. Therefore, an experiment is something that can and should test a theory.

The quantum theory generally cannot predict the result of a single quantum measurement since the quantum theory generally produces only statistical predictions. Aa a result, a single quantum measurement cannot test a statistical prediction of the quantum theory. Therefore a single quantum measurement has no status of an experiment within quantum theory. But what should be considered as an experiment there?

An experiment in quantum theory is a measurement over an ensemble of systems prepared in the same initial state. This idea was thoroughly developed in the school of Dmitry Blokhintsev \cite{Blokhitsev1968} and is known as ``Moscow interpretation of quantum mechanics''\footnote{\texttt{https://www.phys.msu.ru/rus/about/sovphys/ISSUES-2007/1(54)-2007/54-7/}.}. The ensemble is potentially as large as you want collection of systems with the same state. In the measurement over the ensemble the statistical predictions of quantum theory can be checked with any accuracy. Moreover, it is possible to apply any mutual-additive (and mutual-excluding) quantum measurements to the same ensemble of states, that allows to reconstruct the structure of the wave function of the system with any accuracy in all details. Therefore, the wave function of a quantum system is an operationally defined quantity within the formalism of quantum theory.

Here is a simplest example. Consider one-dimensional wave function $\psi(x) = \rho(x)\exp[i\varphi(x)],\ \rho(x) \ge 0$. Using an ensemble of systems in the state $\psi(x)$ we can separately measure probability distributions for coordinate $x$ and momentum $p$, $X(x)$ and $P(p)$ respectively. Then we obtain immediately $\rho(x) = \sqrt{X(x)}$ and the equation
\begin{equation}
 \frac{1}{2\pi\hbar}\int\int dx\,dx'\,\sqrt{X(x)X(x')}e^{\frac{i}{\hbar}p(x - x')}\,e^{-i[\varphi(x) - \varphi(x')]} = P(p)
 \label{eq:XP}
\end{equation}
to obtain $\varphi(x)$. Therefore, the wave function $\psi(x)$ is defined completely. Eq.~(\ref{eq:XP}) determines the phase $\varphi(x)$ up to an additive constant, as expected.

\section{To the nature of mathematics and the mathematical reality}
\label{Mathematics}

The problems related to the nature of mathematics are very complex and extensive, and only a very brief sketch of the author's position is presented here for the possibility of later references.

Consider, for example, the trillionth decimal expansion digit of the number $\sqrt{4711}$. Nobody knows it now, it is not written down anywhere, it simply does not exist \emph{in our material word}. However, if different people start calculating this digit, they will obtain the same result. Why? Because this result objectively \emph{existed} before anyone started to calculate it. But where and how did it exist? This object excised in the objective mathematical reality. Mathematical reality exists objectively, but not in space-time like matter and energy. To exist objectively does not mean to exist necessarily in space and time.

The existence of an objective world of mathematical forms is not a metaphysical statement or philosophical position, since the objective existence of mathematical forms is falsifiable in Popper's sense. Actually, it is enough to present two different correct calculations of the same mathematical object with different results, and the objective existence of this mathematical form and the mathematical reality as a whole will be disproved (falsified). An objection may arise: But the consistent nature of mathematics guarantees that the result will be the same if we solve the same mathematical problem! The reply is: G\"{o}del's second incompleteness theorem says that if mathematics is indeed consistent, then it is impossible to prove its consistence. We principally cannot be absolutely sure of the consistency of mathematics, so comparing the results of different calculations of the same object is always non-trivial. As was written by Nicolas Bourbaki, confidence in the consistency of mathematics is based only on our experience, and on nothing else \cite[p.13]{Bourbaki1968}.

Let us now turn to an overview of the philosophical problems posed by modern physics.

\section{Cosmic variance and the meaning of the theoretical cosmology}
\label{COSMOLOGY}

The standard cosmological model $\Lambda$CDM describes all cosmological observations very well using only 6 free parameters (see for example, \cite[Fig.~1]{PLANCK2020}). The $\Lambda$CDM model predicts among other things probability distributions and expected magnitudes of perturbations of various cosmological fields: temperature, density of matter etc. The predicted probability distributions are distributions over \emph{an ensemble of universes}, or, with additional assumption of homogeneity and isotropy of perturbation fields, they are probability distributions over an ensemble of visible universes. It is in the nature of the theory of cosmological perturbations. Therefore, we must have access to an infinite ensemble of universes to test the predictions of the $\Lambda$CDM cosmology or any other cosmology model with exhaustive accuracy. But we have access only to a single instance of our own visible universe. So we fundamentally cannot test the predictions of cosmological models with exhaustive accuracy -- this is the \emph{cosmic variance} problem.

The theory of cosmological perturbations is a statistical theory, as well as quantum mechanics. Moreover, the source of cosmological perturbations are irreducible quantum fluctuations of the inflaton field (or fields) at the inflation stage of the history of Universe. The source of statisticality in the theory of cosmological perturbations is fundamentally the same as the source of statisticality in quantum theory.

The quantum theory admits the use of arbitrary large ensembles of quantum systems for exact test of statistical predictions of the theory. However, unlike quantum theory, in the theory of cosmological perturbations only one element of the infinite ensemble of universes, which is described by this theory, is accessible. The consequences of this -- the impossibility of exact verification of predictions of cosmological models -- are absolutely dramatic. We mention here only two examples (there are other ones).

\emph{The quadrupole problem.\/} Why is the amplitude of the quadrupole $(l = 2)$ in the distribution of CMB (Cosmic Microwave Background) temperature so low (see \cite[Fig.~1]{PLANCK2020})? Is it an accident event, or may it be a consequence of an unusual topology of the Universe or something else?

\emph{The $l = 20$ problem.\/} What explains the deep dip in the distribution of CMB temperature perturbations near $l = 20$ (see \cite[Fig.~1]{PLANCK2020})? Is it an accident, or is it a defect of $\Lambda$CDM model?

There is no way to answer these questions and there \emph{never} will be. Interestingly, somewhere in the Universe there are places where for accidental reasons (unusual fluctuation of the inflaton field) the CMB anisotropy has nothing in common with the $\Lambda$CDM predictions. What should the inhabitants of such places in the Universe think?

The summary and questions for the cosmic variance problem are:

$\bullet$ Unlike the rest of all other physics (including quantum theory), theoretical cosmology, especially the theory of cosmological perturbations, cannot be precisely verified by experiment due to the internal structure of this statistical theory.

$\bullet$ What then is the epistemological status of cosmology?

$\bullet$ Can cosmology be regarded as a true empirical science?

$\bullet$ Does this situation mean the limit of empirical cognition?

\section{The epistemological status of the Multiverse and ``other universes''}
\label{MULTIVERSE}

The inflation model predicts CMB anisotropy, and, within the standard $\Lambda$CDM model, quantitatively describes CMB anisotropy and other observations very well. CMB anisotropy is the imprinting of irreducible quantum fluctuations of the inflaton field into the picture in our sky. Therefore, without the idea of quantum fluctuations of the inflaton field, what we see in the sky cannot be explained and understood. However, the same irreducible quantum fluctuations, which lead to the visible picture of CMB anisotropy, lead to another prediction: the inflation process generates not only one (our) Universe, but also many other ``local'' universes, which may or may not be similar to our Universe \cite[Ch.~10]{Linde1990}. This multitude of other universes is called the Multiverse. Therefore, it is impossible (very difficult?) to explain the observed CMB anisotropy and understand all other observed phenomena without simultaneously predicting the existence of the Multiverse.

However, each local universe of the Multiverse is completely unreachable for us, since it is separated from us by a space-like interval. We fundamentally have no empirical way to directly test the existence of other universes of the Multiverse, unless there exist some passable ``bridges'' between local universes like wormholes. The questions arises in relation with this strange situation:

$\bullet$ Given that all other universes of the Multiverse lie beyond the reach of direct empirical methods, what is the epistemological status of these objects?

$\bullet$ Should we consider the other universes of the Multiverse to be only objects of mathematical reality (see section~\ref{Mathematics}) arising in the context of the theory of eternal chaotic inflation or they have some special (indirect) empirical status associated with Multiverse being an inherent part of a successful predictive theory?

\section{Operational status of quantum macrostates}
\label{MACROSTATES}

It is supposed in quantum theory that for any quantum state there is a possibility to create an ensemble of any size. In other words, it is supposed the procedure of preparation of a quantum state to be reproducible. Due to this all quantum probabilities and the very notion of quantum state acquire a clear operational (ensemble) meaning (see section~\ref{OperationalQM}).

Macroobjects consist of quantum microsystems and, it would seem, should be quantum objects as well. However (generally speaking) the decoherence time of quantum states of macroobjects is so small that it is fundamentally impossible either to prepare an ensemble of systems in a given state and, even more, to make a measurement over the system. For example, decoherence time of a 10\,$\mu$m dust particle is for temperature 300\,K in 1\,atm air $10^{-31}$\,sec; for temperature 300\,K in absolute vacuum -- $10^{-11}$\,sec \cite{Joos1996}. The question arises: what is the meaning of the statement that a macroobject has a quantum state if this quantum state does not lead to any operationally definable characteristics? It looks that the only type of states of macrosystems for which a reproducible ensemble can be prepared are statistical mixtures indistinguishable from classical probability distributions (the density matrix is strictly diagonal). Should we assume that macrosystems can be characterized by classical states only?

However, is it actually correct that any macrostates are operationally indefinable? No, it is incorrect. For example, there are macrostates separated from the environment and protected from decoherence by an energy gap -- superfluids or superconductors. There are also other ways of protection: topological protection, quantum correction codes in quantum computing. But generally there is no protection against decoherence, therefore decoherence is very strong and operationally-defined quantum description is impossible. The objection is possible: Let us isolate the macrosystem from environment completely. We can use a chamber with walls at absolute zero temperature + absolute vacuum + protection from all radiation, including neutrinos. But this does not solve the problem in the general case, since, first, not all macrosystems make even sense under such deep isolation conditions and, second, it is not always possible to reproducibly prepare the initial state of the macroscopic system even if there is isolation of the system from the environment due to an energy gap or other methods. The following are two important examples.

The first example is about quantum information in biological objects. Quantum informatics allowed to create the first working prototypes of quantum computers. For quantum computers the conditions of quantum coherence conservation and reproducibility of initial state preparation are fulfilled. The question is: Could similar quantum modes of information processing play a role in the brain or even just in any living cell \cite{Penrose1994}? However, there is a problem with this question itself. Even if there are quantum modes in a neuron or in a living cell, sufficiently isolated from the environment (it is, in principle, not impossible), we cannot transfer a living cell to a given quantum initial state in a reproducible way. Therefore it is impossible to create an ensemble of quantum states for a living cell, therefore quantum modes of information processing of a living cell can not have operational sense.

The second example refers to the quantum state of the entire universe. CMB anisotropy is defined by quantum fluctuations of the inflaton field of the scale of the visible event horizon of the Universe. To describe the CMB anisotropy the use of quantum states of the inflaton field of the scale of the visible Universe looks inevitable. However, we observe only a single instance of the Universe, so it is fundamentally impossible to create an ensemble of quantum states for the visible part of the Universe to study it. The quantum state for the visible part of the Universe is operationally indefinable, but we must use this concept for prediction of CMB anisotropy. The situation is very similar to the origin of the phenomenon of cosmic variance (section~\ref{COSMOLOGY}) -- in fact it is the same cosmic variance problem, but translated into the language of quantum ensembles. The problem of cosmic variance, the problem of the quantum state of the Universe and the problem of operational indefinability of macroscopic quantum states are intimately related to each other.

We look, that there is an operationally definable part of quantum theory, where all predicted probabilities and the very notion of quantum state has a well-defined ensemble operational sense (section~\ref{OperationalQM}). At the same time, there are a lot of situations in which it looks inevitably to use operationally undefined quantum probabilities or quantum states. The questions arise:

$\bullet$ Is it true that we have two different quantum theories (ensemble quantum theory and ``Bayesian'' quantum theory)?

$\bullet$ Can the ``Bayesian'' version of quantum theory be considered as a true part of normal empirical science?

\section{The final theory and the meaning of physical reality}
\label{FinalTheory}

Physical theories are represented by mathematical models, and mathematical models in their nature are consistent mathematical systems belonging to the objective world of mathematical forms (section~\ref{Mathematics}). All accepted and confirmed physical theories (Standard Model, $\Lambda$CDM-cosmology, etc.) are considered as approximate descriptions of reality, none of them claiming to be ``complete'' or ``final''. It is assumed that a deeper description of physical reality is possible, from which the existing theories may be derived as approximations or limiting cases.

The question arises: is there a limit to the refinement of physical theories in depth? It’s unknown now. However, it is widely believed that such a ``regression to infinity'' is impossible. In particular, the limit may be related to the Planck scale of energies, distances and times. If ``regression to infinity'' is impossible, then there must be a \emph{final theory} that provides an exhaustive description of physical reality at the deepest level and does not allow for refinements. All other physical theories must be deducible from the final theory as some emergent phenomenology, or, in other words, they all can be reduced to the final theory. Hence another name for the final theory is the \emph{theory of everything}. The search for a final theory is actively pursued. It is supposed that the final theory is some form of quantum gravity plus the unification of gravitation with all other interactions into a single united theory.

If the final theory really exists and admits a mathematical description\footnote{Actually it is not quite clear what may means that the final theory exists, but does not admit mathematical form. This question is not trivial, since we do not know why all our theories succeed in giving mathematical form.}, there is nothing outside this mathematical description. Therefore one of the solution of the final theory turns out to be identical to the physical reality it represents. Unlike all existing theories, which produce approximate mathematical models, solutions of final theory are not models of anything, but is identical with reality they describe. The question arises: what does this mean? There are a number of options to answer, for example:

1. The physical reality is nothing more than a consistent mathematical system supported by the final theory -- one of the exact solutions of the theory.

2. The final theory is a synthetic object of a new type, which is neither a physical reality nor a mathematical system, but it disintegrates into physical reality and a set of usual mathematical models of physics in the ``low-energy limit''.

Max Tegmark was one of the first who clearly posed this question \cite{Tegmark2014}. According to Max Tegmark the answer is obvious, and it is the answer number 1. \emph{In this case} we have the following implication: If the final theory exists, then we live in a ``mathematical matrix'' and we are ourselves mathematical objects. This is \emph{not} a hypothesis, it is a simple logical conclusion from the existence of the final theory, plus answer number 1. But the question remains:

$\bullet$ Is the option number 1 really obvious, are answers more similar option number 2 possible, are any radically other interpretations of physical reality within the final theory possible?

\section{Criticism of superstring theory (M-theory) from the positions of the philosophy of the final theory}
\label{Strings}

In the general relativity, space and time appear to be dynamical objects, and like any dynamical objects, must be subject to the quantum behavior. It follows from quantum theory that space-time must fluctuate very strongly on the Planck scales of length, time and energy. There is no smooth space-time on Planck scale. Therefore, to describe space-time on Planck scale we need a quantum theory of space-time (quantum gravity). Absence of smooth space-time on Planck scale is the initial prerequisite of any quantum theory of gravity.

Superstring theory is one of the candidates for quantum theories of gravity. At the same time, superstring theory is developed as a quantum theory of motion of one-dimensional Planck-scale objects (strings) in smooth space-time. The question immediately arises: What is the smooth background space-time, in which the motion of \emph{Planck-size} strings is considered, given the fact that any quantum gravity theory must start from absence of smooth space-time on Planck scale? One would expect any monograph on string theory to begin with a discussion of what this smooth space-time background means, given that there can be no physical space-time on the Planck scale. However, known to me monographs (some of them are \cite{Green1987,Kaku1988,Brink1988,Deligne1999,Becker2007,Zwiebach2009}) do not discusses this issue at the very beginning or anywhere else.

The smooth ``space-time'' of the string theory background actually cannot relate to the physical space-time. It is by construction a purely abstract background, in which the dynamics of purely formal objects -- strings -- is purely formally considered. It is not defined \emph{a priori} what all this can have to do with real space-time. The smooth background of string theory is much more similar to isotopic spaces for describing internal degrees of freedom of elementary particles than to physical space-time. Rather it looks that this background space somehow describes something like the internal degrees of freedom of a ``quantum of space-time'', just as isotopic spaces describe the internal degrees of freedom of elementary particles.

It is known that a consistent superstring theory can be constructed only in the background space-time of dimension 10 or 11 (the last is for M-theory). On this basis string theorists very often say (especially in popular science publications) that ``our space is 10-dimensional or 11-dimensional, but some space dimensions are compactified on Calabi-Yau manifolds of Planck scale, so these dimensions are not visible to us''. Actually there is no compactification of our space on Planck-size manifolds and there cannot be, because on Planck scales there is no space-time at all, space is not a smooth differentiable manifold on Planck scale, so there is no notion of space dimension and there is simply nothing to compactify. This is directly seen in such quantum theories of gravity as loop quantum gravity or the theory of causal sets where smooth space-time is an emergent phenomenon at large distances only, but the same must take place also in the general case. The string theorists' claims about 10 or 11 dimensions refer actually only the formal mathematical structure of the background of string theory, not to physical space-time. How our physical space-time is related to string theory is a very difficult question.

Now return back to the philosophy of the final theory. In the final theory the most fundamental level of physical reality is some mathematical structure identical to the physical reality at the deepest level. All other physics is obtained from this structure in some emergent and unknown beforehand way. It is important that this mathematical structure itself does not have to have any characteristics directly corresponding to emergent physical concepts of higher level. The structure of string theory is very similar to the expected structure of the final theory. There is nothing wrong in the fact that string theory is built in an abstract smooth background space, which has no direct relation to physical space-time. It is only necessary to realize that the abstract background of string theory is not the physical space-time. Quantum dynamics of strings in formal smooth background can be those consistent mathematical structure, which defines the fundamental physical reality. Physical space-time must be obtained from here in some emergent way, but this way is not predefined in advance and may not be simple and straightforward.

String theory is not the only candidate for a final theory (quantum gravity). Other candidates are: different forms of loop quantum gravity and spin foam models \cite{Rovelli2014}, different forms of causal set theory \cite{Surya2019}, etc.  All of them have some features of the final theory in that they are based on formal mathematical structures rather far from observations, and these mathematical structures often have even an abstract combinatorial nature, not space-time or field nature, as in the string theory. We do not know whether is there a ``correct candidate'' among the contemporary applicants for the role of the true final theory.

%

\begin{thebibliography}{10}
\def\selectlanguageifdefined#1{
\expandafter\ifx\csname date#1\endcsname\relax
\else\selectlanguage{#1}\fi}
\providecommand*{\href}[2]{{\small #2}}
\providecommand*{\url}[1]{{\small #1}}
\providecommand*{\BibUrl}[1]{\url{#1}}
\providecommand{\BibAnnote}[1]{}
\providecommand*{\BibEmph}[1]{\emph{#1}}
\ProvideTextCommandDefault{\cyrdash}{\hbox to.8em{--\hss--}}
\providecommand*{\BibDash}{\ifdim\lastskip>0pt\unskip\nobreak\hskip.2em\fi
\cyrdash\hskip.2em\ignorespaces}

\bibitem{Blokhitsev1968}
\selectlanguageifdefined{english}
\BibEmph{Blokhintsev D.I.} {The Philosophy of Quantum Mechanics}. \BibDash
\newblock D.Reidel, 1968.

\bibitem{Bourbaki1968}
\selectlanguageifdefined{english}
\BibEmph{Bourbaki N.} {Elements of mathematics. Theory of sets.} \BibDash
\newblock Springer, 1968.

\bibitem{PLANCK2020}
\selectlanguageifdefined{english}
\BibEmph{Aghanim N., Akrami Y., M. A., {et al. (Planck collaboration)}.}
  {Planck 2018 results. VI. Cosmological parameters}~//
  \href{http://dx.doi.org/10.1051/0004-6361/201833910}{Astronomy and
  Astrophysics}. \BibDash
\newblock 2020. \BibDash
\newblock V. 641. \BibDash
\newblock P.~A6. \BibDash
\newblock arXiv:1807.06209.

\bibitem{Linde1990}
\selectlanguageifdefined{english}
\BibEmph{Linde A.} {Particle physics and inflationary cosmology}. \BibDash
\newblock CRC Press, 1990.

\bibitem{Joos1996}
\selectlanguageifdefined{english}
\BibEmph{Joos E.} {Decoherence Through Interaction with the Environment}~//
  {Decoherence and the Appearance of a Classical World in Quantum Theory}.
  \BibDash
\newblock Springer, 1996. \BibDash
\newblock P.~37--146.

\bibitem{Penrose1994}
\selectlanguageifdefined{english}
\BibEmph{Penrose R.} {Shadows of the Mind: A Search for the Missing Science of
  Consciousness}. \BibDash
\newblock Oxford University Press, 1994.

\bibitem{Tegmark2014}
\selectlanguageifdefined{english}
\BibEmph{Tegmark M.} {Our Mathematical Universe: My Quest for the Ultimate
  Nature of Reality}. \BibDash
\newblock Knopf, 2014.

\bibitem{Green1987}
\selectlanguageifdefined{english}
\BibEmph{Green M.B., Schwarz J.H., Witten E.} {Superstring Theory. Volumes 1
  and 2}. \BibDash
\newblock CAMBRIDGE UNIVERSITY PRESS, 1987.

\bibitem{Kaku1988}
\selectlanguageifdefined{english}
\BibEmph{Kaku M.} {Introduction to Superstrings}. \BibDash
\newblock Springer-Verlag, 1988.

\bibitem{Brink1988}
\selectlanguageifdefined{english}
\BibEmph{Brink L., Henneaux M.} {Principles of String Theory}. \BibDash
\newblock PLENUM PRESS, 1988.

\bibitem{Deligne1999}
\selectlanguageifdefined{english}
{Quantum Fields and Strings: A Course for Mathematicians. Volume 2.}~/
  P.~Deligne, D.~Kazhdan, P~Etingof, J.~W.~Morgan, D.~S.~Freed, D.~R.~Morrison,
  L.~C.~Jeffrey, E.~Witten. \BibDash
\newblock American Mathematical Society, Institute for Advanced Study, 1999.

\bibitem{Becker2007}
\selectlanguageifdefined{english}
\BibEmph{Becker K., Becker M., Schwarz J.H.} {String Theory and M-Theory. A
  Modern Introduction}. \BibDash
\newblock CAMBRIDGE UNIVERSITY PRESS, 2007.

\bibitem{Zwiebach2009}
\selectlanguageifdefined{english}
\BibEmph{Zwiebach B.} {A First Course in String Theory. Second Edition}.
  \BibDash
\newblock CAMBRIDGE UNIVERSITY PRESS, 2009.

\bibitem{Rovelli2014}
\selectlanguageifdefined{english}
\BibEmph{Rovelli C., Vidotto F.} {An elementary introduction to Quantum Gravity
  and Spinfoam Theory}. \BibDash
\newblock CAMBRIDGE UNIVERSITY PRESS, 2014.

\bibitem{Surya2019}
\selectlanguageifdefined{english}
\BibEmph{Surya S.} {The causal set approach to quantum gravity}~//
  \href{http://dx.doi.org/10.1007/s41114-019-0023-1}{Living Reviews in
  Relativity}. \BibDash
\newblock 2019. \BibDash
\newblock V.~22. \BibDash
\newblock P.~5. \BibDash
\newblock arXiv:1903.11544.

\end{thebibliography}

\end{document}